\documentclass[%
reprint,
 amsmath,amssymb,
aps,
pra,
]{revtex4-2}

\usepackage{graphicx}
\usepackage{dcolumn}
\usepackage{bm}
\usepackage{import}
\usepackage{braket}
\usepackage{color}
\usepackage{transparent}
\usepackage{nicefrac}
\usepackage{siunitx}
\usepackage{upgreek}
\usepackage{tabularx}
\usepackage{bigstrut}
\usepackage{ctable}
\usepackage{booktabs}

\usepackage{hyperref}
\hypersetup{
	colorlinks,
	linkcolor={red!50!black},
	citecolor={blue!50!black},
	urlcolor={blue!80!black}
}

\newcommand{\bk}{\mathbf{k}}

\begin{document}

\preprint{APS/123-QED}

\title{Doppler-free three-photon coherence effects in mercury vapor}

\author{Benjamin Rein}
\author{Jochen Schmitt}%
\author{Martin R. Sturm}
\author{Reinhold Walser}
\author{Thomas Walther}
\email{thomas.walther@physik.tu-darmstadt.de}
\affiliation{Institute for Applied Physics, TU Darmstadt, 64289 Darmstadt, Germany.}%



\date{\today}

\begin{abstract}
Doppler-free three-photon coherence effects at a wavelength of \SI{253.7}{nm} have been observed in thermal mercury vapor. The experimental results are compared to simulations based on a detailed theoretical model reproducing the measured effects. Based on these results, we show by simulations that amplification without inversion in our setup is feasible, paving the way towards lasing without inversion in mercury.
\begin{description}
\item[PACS numbers]
\end{description}
\end{abstract}
\pacs{42.50.Gy}
\pacs{42.50.-p}
\pacs{32.80.Qk}
%
\maketitle


\section{\label{sec:introduction}Introduction}
The development of continuous-wave (cw) lasers in the ultraviolet (UV) or even vacuum ultraviolet (VUV) wavelength region is still a challenging task as the pump power requirements to induce a population inversion on the laser transition scales at least with the laser frequency $\omega^4$~\cite{Mompart2000} being the decisive obstacle of building conventional cw lasers in this wavelength regime~\cite{Mandel1993,Lukin1996}. Common methods to generate cw laser radiation in the UV or even VUV are based on nonlinear processes such as second- and fourth-harmonic generation~\cite{Sudmeyer2008} as well as four-wave mixing~\cite{Kolbe2012}. However, these techniques are limited by the availability of materials transparent in the VUV and require a large fundamental power due to their nonlinear character. Furthermore, to achieve an increasingly shorter wavelength with these techniques, the fundamental or coupling lasers also have to be at an increasingly shorter wavelength.\par
A completely different approach is represented by lasing without inversion (LWI) first proposed in~\cite{Kocharovskaia1988,Scully1989,Harris1989,Imamoglu1989}. The main idea of LWI is to suppress the absorption of coherent radiation on the lasing transition by using quantum interference effects such as electromagnetically induced transparency (EIT) \cite{harris91,harris91a,harris97,sulzbach19} or coherent population trapping (CPT). In contrast to nonlinear techniques where the medium mediates the energy transfer of the fundamental lasers to coherent radiation at a shorter wavelength, in LWI the energy is transformed from incoherent to coherent radiation in the medium comparable to conventional lasers.\par
The feasibility of cw LWI has already been demonstrated in three-level schemes for rubidium~\cite{Zibrov1995} and sodium~\cite{Fry1993,Padmabandu1996}, but until now there has been no LWI scheme where the laser transition had a significantly shorter wavelength than the coupling lasers. The main obstacle here is the Doppler-broadening reducing the overall gain for LWI when the wavelength of the coupling laser and the actual laser transition are separated in energy~\cite{Lukin1996}. A four-level scheme in mercury first proposed in ref.~\cite{Fry2000a} makes it possible to cancel the Doppler-effect, preventing the gain spike from being washed out, enabling LWI at \SI{253.7}{nm} while the shortest wavelength of the coupling lasers is at \SI{435.8}{nm}.\par
We report on the experimental implementation of this four-level scheme in mercury and present the first measurements of a Doppler-free three-photon coherence in thermal mercury vapor, forming the basis of observing LWI at a short wavelength in the UV. Furthermore, we estimate the incoherent pump power necessary for achieving amplification without inversion (AWI) through measurements and detailed model calculations~\cite{Sturm2014}.\par
%

%
%
%
\section{Overview of LWI in mercury}
\label{sec:theory}
Disturbing a coherently driven three-level system (e.g. an EIT-system) basically leads to an attenuation of the coherent effect caused by the occurrence of decoherence~\cite{Scully1999,Lukin1999}. Nevertheless, it is possible to couple a coherently driven three-level system to a fourth level through weak laser radiation. This targeted interference can lead to a splitting of the dark-state of the three-level system, the so called interacting dark resonances or double-dark-states (DDKS)~\cite{Lukin1999,Ye2002b,Wei2007a}.
\begin{figure}[h]
	\centering
	\includegraphics[width=\linewidth]{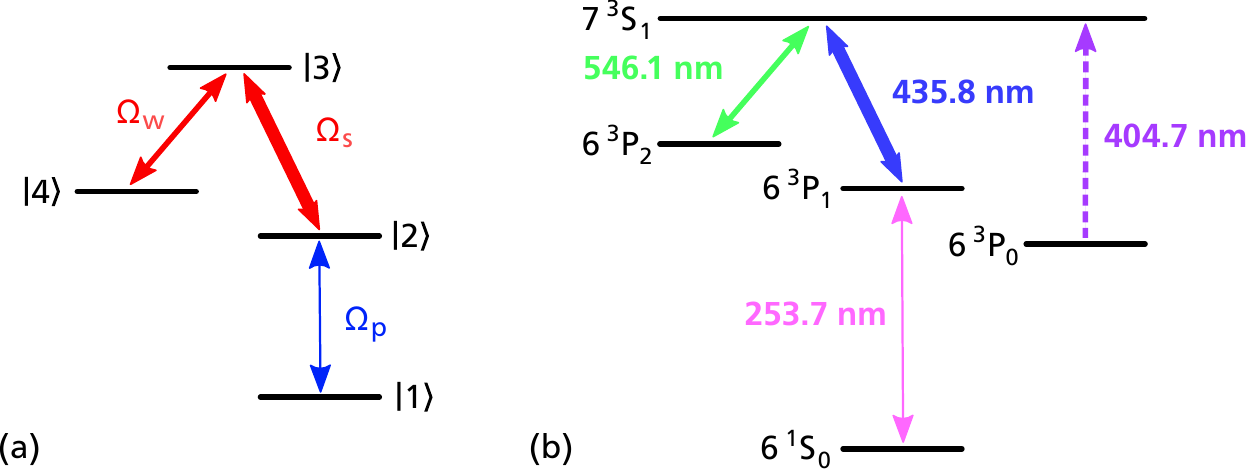}
	\caption{(a) Four-level scheme with LWI laser transition $\Omega_{p}$, strong coupling laser $\Omega_{s}$ and weak coupling laser $\Omega_{w}$. (b) Level structure of mercury. The metastable $6{^3}P_0$ state is not part of the LWI scheme but has to be taken into account as population can be trapped in this level.}
	\label{fig:LWI-Schema-Vergleich}
\end{figure}%

Such a four-level scheme is depicted in Fig.~\ref{fig:LWI-Schema-Vergleich} (a) where the levels $\ket{1}, \ket{2}$ and $\ket{3}$ correspond to an EIT ladder system, where the transition $\ket{2} \leftrightarrow \ket{3}$ is coherently coupled by a strong laser with the Rabi frequency $\Omega_s$ and the transition $\ket{1} \leftrightarrow \ket{2}$ is probed by a laser with Rabi frequency $\Omega_p$. In addition, level $\ket{3}$ is coherently coupled to the metastable level $\ket{4}$ through a weak laser with Rabi frequency $\Omega_w$. %

To understand the origin of the DDKS, we write down the Hamiltonian of the system in the basis of the bare atomic states $\{ \ket{2}, \ket{1}, \ket{3}, \ket{4} \}$ in dipole- and rotating wave approximation~\cite{Sturm2014}
\begin{align}
\label{equ:4NS-Hamiltonian}
H = -\hbar \begin{pmatrix}
-\Delta_{p} & \Omega_{p} & 0 & 0 \\
\Omega_{p}^* & 0 & \Omega_{s} & 0 \\
0 & \Omega^*_{s} & \Delta_{s} & \Omega^*_{w}  \\
0 & 0 & \Omega_{w}  & \Delta_{s} - \Delta_{w} 
\end{pmatrix}
\end{align}
where $\Delta_{p,s,w}=\omega_{12,23,34}-\omega_{p,s,w}$ are the detunings of the lasers from the respective atomic transitions $\omega_{ij}$. The dressed-state picture gives a better insight into the appearance of the DDKS. For this purpose, the Hamiltonian matrix is diagonalized and solved for the eigenstates:
\begin{align}
\ket{0} &= \ket{4} - \frac{\Omega^*_w}{\Omega^*_s} \ket{2} \\
\ket{\pm} &= \frac{1}{\sqrt{2}} \left( \ket{2} \mp \frac{|\Omega_s|}{\Omega_s} \ket{3} + \frac{\Omega_w}{\Omega_s} \ket{4} \right),
\end{align}
with all detunings $\Delta_{p,s,w} = 0$ and for a weak probe beam such that $\Omega_p \rightarrow 0$ and first order contributions from $\Omega_w$. Without the weak coupling laser, $\Omega_w = 0$, the eigenstates $\ket{\pm}$ correspond to the usual Autler-Townes components of a three-level ladder EIT-system and $\ket{0}$  to the bare atomic state $\ket{4}$, respectively. When the weak coupling laser is switched on, the eigenstate $\ket{0}$ contains an admixture of the bare-states $\ket{2}$ and $\ket{4}$ and therefore couples to the ground state $\ket{1}$. In the bare-state picture this corresponds to a three-photon resonance and leads to a very sharp absorption feature~\cite{Lukin1999,Sturm2014}.\par
The Bloch equations derived in the appendix allow for a quantitative investigation of the three-photon resonance. Here, the decay rates $\Gamma_{12}$, $\Gamma_{23}$, and $\Gamma_{34}$ of the respective atomic transitions are incorporated. The central result is the system's linear response to the probe field $\Omega_p$ in the stationary limit, the linear susceptibility
\begin{equation}
	\chi^{(1)} =\chi' + i \chi'' = \kappa
	\frac{i \Delta_{3} \gamma_{1} - |\Omega_w|^2}
	{\Delta_{3} \left( \gamma_1 \gamma_2 + |\Omega_s|^2\right) + i \gamma_2 |\Omega_w|^2},
\end{equation}
with the three-photon detuning $\Delta_{3} = \Delta_p + \Delta_s - \Delta_w$, and the two complex decay rates
$\gamma_2=\Gamma_{12}/2-i \Delta_p$ as well as
$\gamma_1 = (\Gamma_{23}+\Gamma_{34})/2 - i (\Delta_p + \Delta_s)$. 
The dimensionless prefactor $\kappa = \mathcal{N} \left| d_{21} \right|^2 / (\hbar \; \epsilon_0)$ can be calculated from the atomic density $\mathcal{N}$ and the atomic dipole matrix element $d_{21}$.
As usual, the real part $\chi'$ describes dispersion, whereas the imaginary part $\chi''$ corresponds to absorption ($\chi'' > 0$) or gain ($\chi'' < 0$).\par
\begin{figure}[tb]
	\centering
	\includegraphics[width=\linewidth]{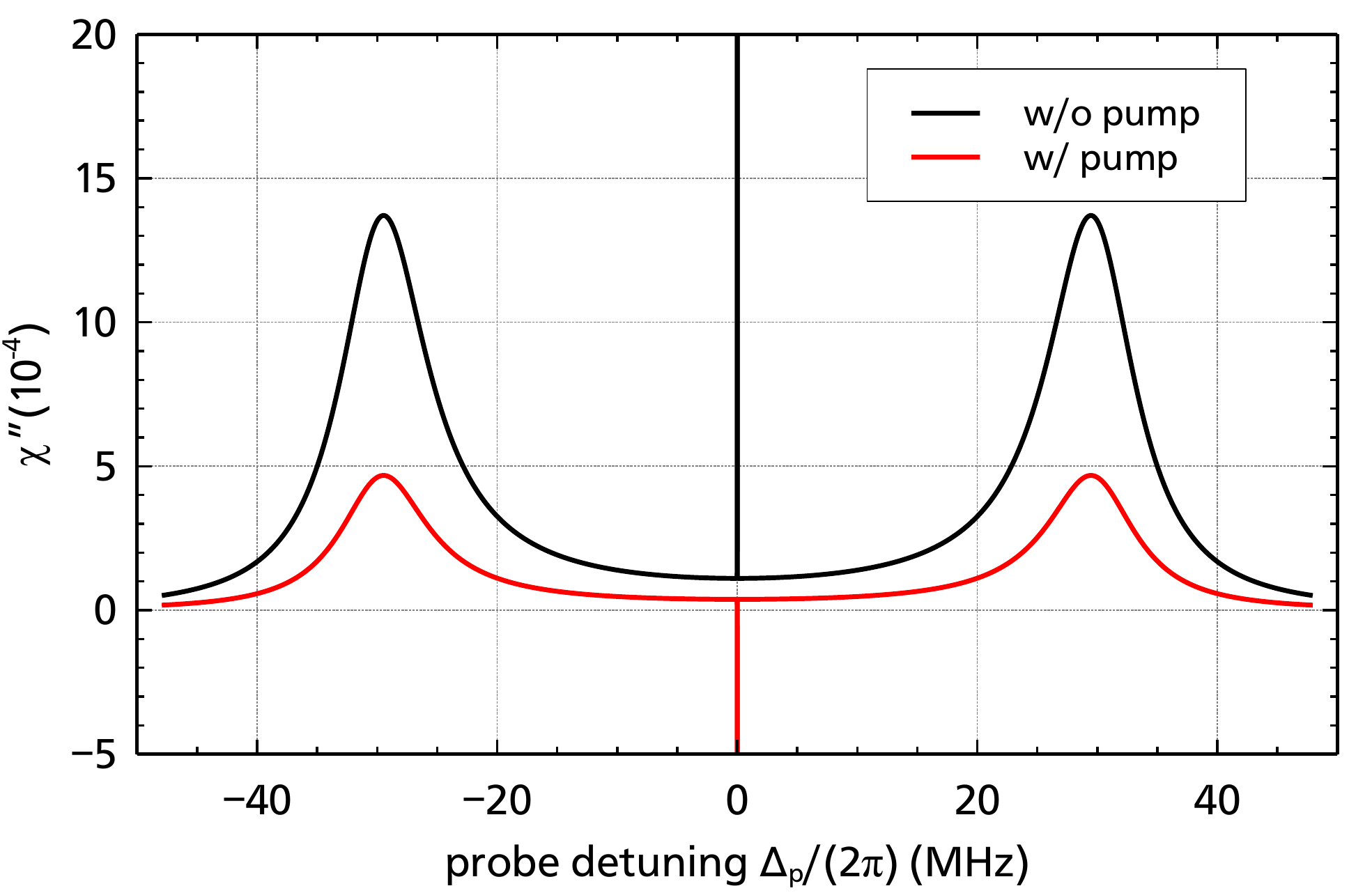}
	\caption{Imaginary part of the probe transition's linear susceptibility $\chi''$ versus the detuning of the probe laser $\Delta_p$ for $\Omega_s = 2 \pi \cdot 30 \; \text{MHz}$, $\Omega_w = 2 \pi \cdot 0.5 \; \text{MHz}$, $\Delta_s = \Delta_w = 0$, $\Gamma_{12}=2 \pi \cdot 1.27 \; \text{MHz}$, $\Gamma_{23}=2 \pi \cdot 8.86 \; \text{MHz}$, $\Gamma_{34}=2 \pi \cdot 7.75 \; \text{MHz}$, and $\kappa = 7.57 \cdot 10^{4} \; \text{s}^{-1}$. For the black graph no pumping on $1 \leftrightarrow 2$ transition was applied whereas for the red graph the incoherent pumprate was $r = 5 \cdot 10^3 \; \text{s}^{-1}$.}
	\label{fig:Susceptibility}
\end{figure}%

To obtain this three-photon resonance even for Doppler-broadened media and wavelength differences of several \si{nm}
between probe and coupling lasers, it is possible to use a specific geometrical orientation of the laser beams as proposed in~\cite{Fry2000a} as long as the $\mathbf{k}$-vectors involved obey the triangle equation. The linear Doppler-shift of the three-photon resonance $\Delta_{3}$ experienced by an atom moving with the velocity $\mathbf{v}$ yields
\begin{align}\label{equ:Doppler-broadening}
\Delta_{3} = 
\Delta_p + \Delta_s - \Delta_w 
- \left( \bk_p + \bk_s - \bk_w \right) \cdot \mathbf{v},
\end{align}
where $\bk_{p,s,w}$ are the wave vectors of the laser beams. If a geometrical orientation of the laser beams can be found such that $  \bk_w =\bk_p + \bk_s$, the Doppler-shift $\Delta_{3 }= 0$, independent of the atom's velocity when the laser beams are in resonance with the atomic transitions ($\Delta_{p,s,w} = 0$). This allows the use of a gas cell  simplifying the experimental realization and allowing for higher atomic densities than in e.g. atomic beams or magneto-optical traps.\par
Mercury has a level structure meeting the requirements of a four-level LWI system as stated above (cf. Fig.~\ref{fig:LWI-Schema-Vergleich} (b)). The LWI laser transition corresponds to the first ground state transition in the UV at \SI{253.7}{nm}. The strong coupling laser has a wavelength of \SI{435.8}{nm} and the weak coupling laser at \SI{546.1}{nm} couples to the metastable $6{^3}P_2$ state with a lifetime in the range of seconds~\cite{Garstang1962}. In order to prevent  population trapping in the metastable  $6{^3}P_0$ state through spontaneous emission from the $7{^3}S_1$ state, an incoherent repump at \SI{404.7}{nm} is necessary. A great benefit of this LWI scheme is that radiation at a wavelength of \SI{253.7}{nm} can be obtained by fourth-harmonic generation so that the spectroscopic properties can be analyzed by employing a weak probe laser at this wavelength. The influence of parameters of the coupling lasers such as linewidth, detuning and power is of particular interest since this is the first four-level LWI system being investigated experimentally. 
\begin{figure}[tb]
	\centering
	\includegraphics[width=0.8\linewidth]{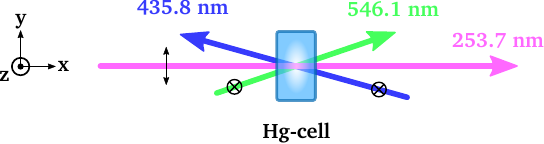}
	\caption{Superposition of the three laser beams in a mercury vapor cell taking into consideration the angles for the Doppler-free configuration so that $\bk_{253.7} + \bk_{435.8} = \bk_{546.1}$ and the polarizations proposed in \cite{Fry2000a}. The length of the three laser beams are scaled according to the absolute value of their wavevectors.}
	\label{fig:LWI-Winkelschema}
\end{figure}%
To satisfy the condition for a Doppler-free three-photon resonance so that $\Delta_{\textrm{3PR}} = 0$, the three laser beams have to be aligned as depicted in Fig.~\ref{fig:LWI-Winkelschema}, where the strong coupling beam is counterpropagating with respect to the probe beam at an angle of \SI{15.3}{\degree} and the weak coupling beam is copropagating at an angle of \SI{19.3}{\degree}. Fig.~\ref{fig:LWI-Winkelschema} also shows the superposition of the laser beams in a mercury vapor cell and their linear polarizations as proposed in \cite{Fry2000a}.\par
Based on the measurements presented in this paper, the theoretical model by Sturm \textit{et. al.}~\cite{Sturm2014} was extended to also include the finite interaction time $t_{\textrm{int}}$ of the mercury atoms with the laser beams. At a temperature of \SI{16}{\degreeCelsius} in the absorption cell, the atoms have a velocity of about \SI{155}{m\per s}. With a beam diameter of \SI{0.84}{mm} for the \SI{253.7}{nm} probe beam, we find  $t_{\textrm{int}} = \SI{5.4}{\micro s}$, defining the upper limit for the coherence lifetime. Furthermore, the incoherent pump is now modelled as a directed pump beam as used in the experiment also taking into account its spectral width ($\textrm{FWHM}_\textrm{pump}$). It is a crucial parameter as the three-photon resonance is essentially Doppler-free. As a consequence, all atoms regardless of their velocity class contribute to the coherence effect. Assuming a directed pump beam, its spectral width determines how many atoms of the Doppler-broadened ensemble can be addressed, which in turn is a measure of the pump's efficiency.\par
\begin{figure}[htb]
	\centering
	\includegraphics[width=\linewidth]{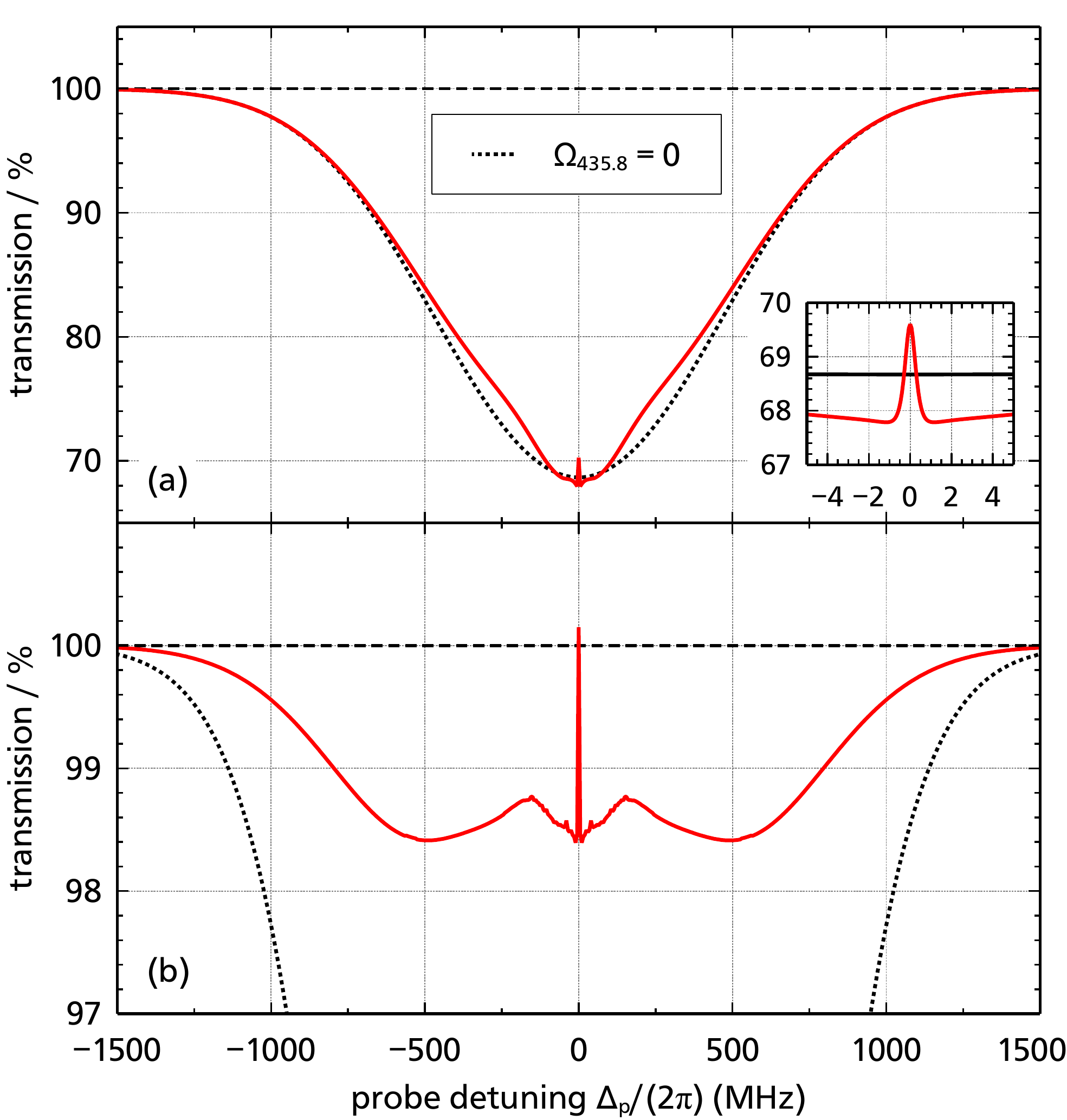}
	\caption{Absorption spectra of the $6^{1}S_{0} \leftrightarrow 6^{3}P_{1}$ transition calculated with the extended theoretical model. The black dotted line shows the undisturbed absorption for comparison. (a) Absorption spectrum with coupling lasers and repumper enabled (red line). The inset shows the Doppler-free three-photon resonance peak. (b) Absorption spectrum with an incoherent pump at \SI{253.7}{nm}, showing AWI for the probe beam (red line).}
	\label{fig:3P-Simulation}
\end{figure}

Fig.~\ref{fig:3P-Simulation} shows absorption spectra of the $6^{1}S_{0} \leftrightarrow 6^{3}P_{1}$ transition calculated with the extended theoretical model with laser parameters as in the experiment (cf. Tab. \ref{tab:Laserparameter}) and an absorption path of \SI{2}{mm}. 
The only difference is the linewidth of the \SI{546.1}{nm} coupling laser which was measured to be \SI{183}{kHz}. For the calculation a linewidth of \SI{60}{kHz} was chosen for a better fit to the measured data. A possible cause of this deviation could be the different noise components as described in~\cite{Ludvigsen1998, Fuhrer2012} and their effect on the coherent stimulation. This will be investigated more closely in the future. Furthermore, the linewidth of the probe laser was also estimated to be \SI{60}{kHz} but could not be verified experimentally. The temperature of the mercury atoms was assumed to be \SI{16}{\degreeCelsius}, resulting in a Doppler-width of \SI{1013}{MHz}, while for the transmission a temperature of \SI{5}{\degreeCelsius} was assumed (for an explanation of this discrepancy, see section~\ref{sec:set-up}). The black line in Fig.~\ref{fig:3P-Simulation} (a) and (b) shows the undisturbed Doppler-broadened absorption spectra. When the two coupling lasers and the incoherent repump are switched on, the spectrum is modified as shown by the red line in Fig.~\ref{fig:3P-Simulation} (a). The inset shows the central \SI{\pm 5}{MHz} of the spectrum where the three-photon resonance peak is visible. The full width at half maximum (FWHM) of this peak is \SI{704}{kHz} which is much less than the natural linewidth of the $6^{1}S_{0} \leftrightarrow 6^{3}P_{1}$ transition with \SI{1.27}{MHz}. With an incoherent pump beam, copropagating with the probe beam at an angle of \SI{5}{\degree}, a pump rate of $2\pi\cdot$\SI{10}{MHz} and a spectral width of \SI{100}{MHz} the three-photon resonance is shifted into the gain region (transmission $>$ \SI{100}{\%}) as shown in Fig.~\ref{fig:3P-Simulation} (b) (red line), demonstrating the possibility of AWI in mercury, which in turn is a necessity for LWI.
%
%
%
\section{Experimental realization}
\label{sec:set-up}
\begin{figure*}[tb]
	\centering
	\includegraphics[width=\linewidth]{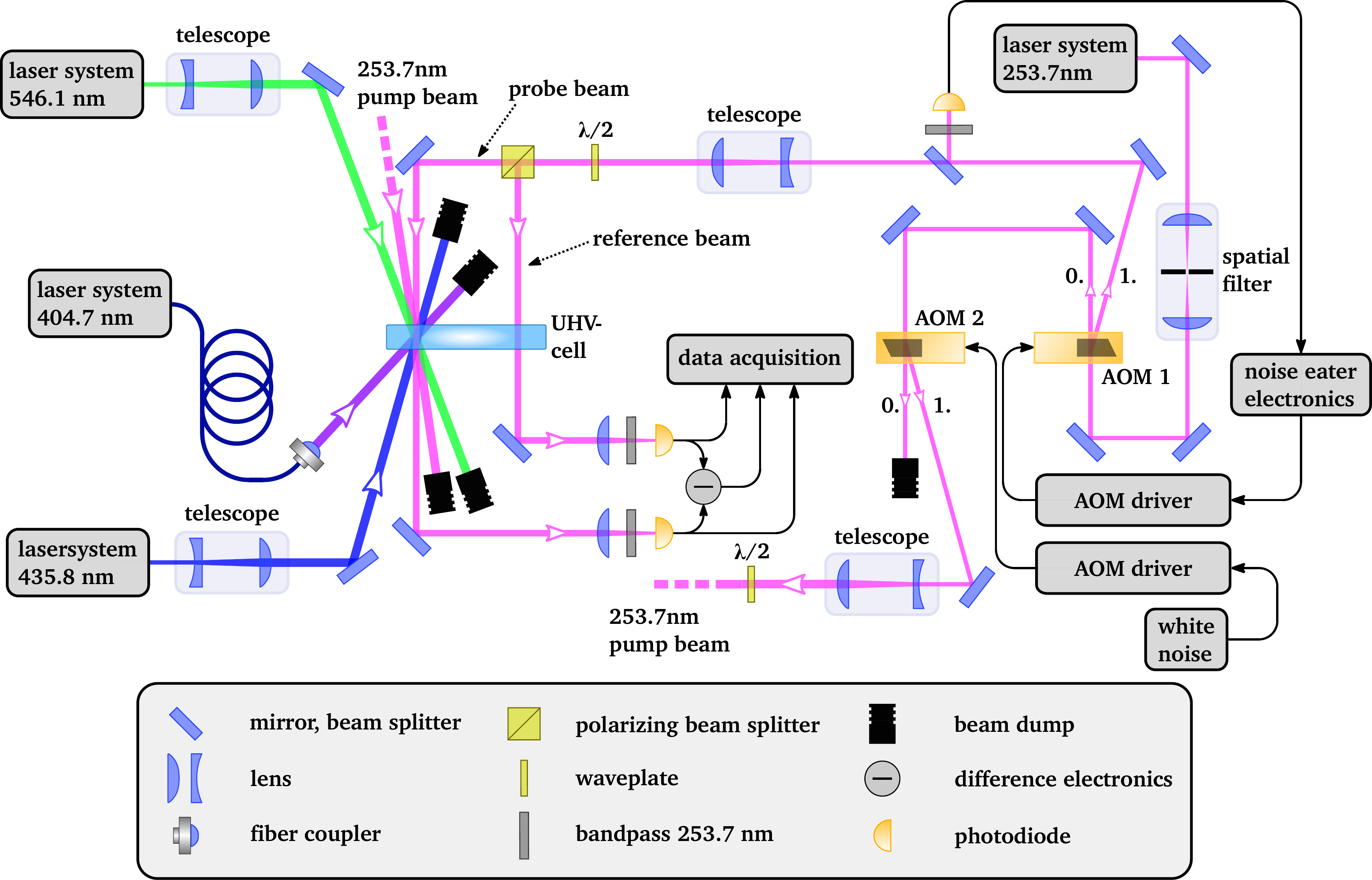}
	\caption{Schematic diagram of the experimental setup. The angles of the laser beams overlapping in the UHV cell are to scale as in the actual experimental implementation. The vacuum chamber connected to the UHV cell is not shown in this figure.}
	\label{fig:UV-Strahlengang}
\end{figure*}

Fig.~\ref{fig:UV-Strahlengang} shows a schematic diagram of the experimental setup.
The key parts are the two coherent coupling lasers at \SI{435.8}{nm} and \SI{546.1}{nm} since their linewidth and frequency stability are important as they directly influence the three-photon resonance. Therefore, all laser systems are based on external cavity diode lasers (ECDL) with typical linewidths below \SI{100}{kHz}~\cite{Wieman1991b}. The radiation at \SI{435.8}{nm} and \SI{546.1}{nm} are generated by amplifying the fundamental radiation of an ECDL by a tapered amplifier followed by nonlinear frequency doubling. Both systems are frequency stabilized utilizing polarization spectroscopy with a peak-to-peak stability better than \SI{495}{kHz}. Details of the \SI{435.8}{nm} laser system can be found in~\cite{Rein2017b}. As the incoherent repump, we use an ECDL at \SI{404.7}{nm} artificially broadened up to \SI{68}{MHz} by directly modulating the laser diode current with white noise and simultaneous frequency stabilization by dichroic spectroscopy~\cite{Rein2017}.\par
In order to verify the theoretical model we use a \SI{253.7}{nm} laser system based on frequency quadrupling, tunable around the $6^{1}S_{0} \leftrightarrow 6^{3}P_{1}$ transition. A spatial filter ensures a Gaussian beam shape of the UV radiation. About \SI{200}{\micro W} are diffracted in the first order of
an acousto-optic modulator (AOM) and power stabilized by a "noise eater" circuit. This beam is split equally into a probe and reference beam, respectively. The residual UV power is sent through a second AOM whose carrier frequency is modulated with white noise with a maximum bandwidth of \SI{10}{MHz} thus providing the pump beam for the $6^{1}S_{0} \leftrightarrow 6^{3}P_{1}$ transition \cite{Elliott1982a}. However, the AOM driver was not capable to transfer the white noise bandwidth onto the carrier frequency and only a frequency broadening of \SI{0.5}{MHz} could be measured with an electric spectrum analyzer.\par
The three laser systems at \SI{404.5}{nm}, \SI{435.8}{nm} and \SI{546.1}{nm} are stabilized to the $^{202}\textrm{Hg}$ isotope since it has the highest abundance with \SI{29.65}{\%}~\cite{Sansonetti2005a}. The probe and reference beam is scanned about \SI{7}{GHz}, centered around the transition of the $^{202}\textrm{Hg}$ isotope. The scan speed is about \SI{9}{s} to ensure that the two doubling stages could follow the frequency scan without introducing any instabilities. The utilized laser parameters such as power, beam diameters and resulting Rabi frequencies are listed in Tab.~\ref{tab:Laserparameter}. %
\begin{table*}
	\caption{Relevant parameters of the four lasers employed for the measurements of the Doppler-free three-photon coherence.}
	\centering
	\begin{ruledtabular}
	\begin{tabular}{lrrrrr}
		Wavelength & Power & \begin{tabular}{@{}l@{}}Beam-\\diameter\end{tabular} & \begin{tabular}{@{}l@{}}Rabifrequency\\$\nicefrac{\Omega}{2\pi}$\end{tabular} & \begin{tabular}{@{}l@{}}Saturation-\\parameter $S_0$\end{tabular} &  Linewidth\\\hline\addlinespace
		\SI{253.7}{nm} & \SI{50}{\micro W} & \SI{0.84}{mm} & \SI{0.21}{MHz} & 0.88 &  \SI{26}{kHz} (for \SI{1014.8}{nm}) \\\addlinespace
		\SI{435.8}{nm} & \SI{170}{mW} & \SI{2}{mm} & \SI{30,8}{MHz} & 385 & \SI{60}{kHz}\\\addlinespace
		\SI{546.1}{nm} & \SI{3.95}{mW} & \SI{2}{mm} & \SI{6.17}{MHz} & 20.2 & \SI{183}{kHz}\\\addlinespace
		\SI{404.7}{nm} & \SI{3.7}{mW} & \SI{2.8}{mm} & \SI{1.80}{MHz} & 9.10 & \SI{52}{MHz}\\
	\end{tabular}
	\end{ruledtabular}
	\label{tab:Laserparameter}
\end{table*}%

All laser beams except for the reference beam are overlapped within an ultra-high vacuum cell (UHV cell) according to the Doppler-free configuration as shown in Fig.~\ref{fig:LWI-Winkelschema}. The reference beam is vertically shifted with respect to the overlap region, experiencing only the undisturbed, Doppler-broadened absorption. Since the UHV cell features a tapered shape, the absorption path can be adjusted between \SI{1.5}{mm} to \SI{6}{mm} and is fixed to \SI{2}{mm} for the measurements presented in this paper. The beam diameters for the probe and reference beam, the two coupling lasers and the repumper are listed in Tab.~\ref{tab:Laserparameter}.\par
As the laser linewidth is a crucial parameter, it was measured using a self-heterodyne setup~\cite{Ludvigsen1998,Fuhrer2012}. While the linewidth was directly determined for the wavelength of \SI{435.8}{nm} and \SI{546.1}{nm} using a \SI{2.06}{\micro s} delay-time by introducing an optical delay by a fiber for part of the beam before impinging both parts on a fast photo diode. For the wavelength of \SI{253.7}{nm} it was determined for the fundamental laser at \SI{1014.8}{nm} and a delay-time of \SI{1.71}{\micro s}. The linewidth of the \SI{404.7}{nm} repump was measured with a high-finesse Fabry-P\'{e}rot interferometer. The measured linewidths are also given in Tab. \ref{tab:Laserparameter}.\par 
The UHV cell is connected to a vacuum chamber including a temperature stabilized mercury reservoir (temperature range between \SI{-40}{\degreeCelsius} to \SI{+30}{\degreeCelsius}) leading to an adjustable optical density of the mercury atoms. During our measurements the reservoir temperature is fixed to \SI{+5}{\degreeCelsius}, resulting in a transmission of the reference beam of about \SI{68}{\%} for a \SI{2}{mm} absorption path. While the optical density can be varied, the actual temperature of the mercury atoms  stays almost constant at \SI{+16}{\degreeCelsius}, independent of the reservoir temperature due to the spatial separation between the UHV cell and the reservoir in the vacuum chamber, resulting in a Doppler-width of about \SI{1013}{MHz}. \par
The probe and reference beams are detected by UV photodiodes (JEC4, Laser Components GmbH) with a spectral sensitivity between \SI{210}{nm} to \SI{380}{nm} and protected from ambient light by additional bandpass filters (Hg01-254-25, Semrock Inc.) with a central wavelength of \SI{250}{nm} and a bandwidth of \SI{15}{nm}. The two photodiodes are connected in series in order to directly measure the difference signal and cancel out residual power fluctuations of the UV beam, which are equal in probe and reference beam leading to a better signal to noise ratio. In addition, it is possible to simultaneously measure the signals of the individual photodiodes.\par
To prevent Zeeman-shifts caused by the Earth's magnetic field, it is compensated by a set of three Helmholtz coils arranged around the overlap region.\par
%
%
%
\section{Measurements of the three-photon coherence}
The influence of the two coupling lasers and the repumper is measured using four configurations as summarized in Tab. \ref{tab:3PR-mconfig}. %
\begin{table}
	\caption{Measurement configurations used to investigate the influence of the individual lasers.}
	\centering
	\begin{ruledtabular}
	\begin{tabular}{lcccc}
		\begin{tabular}{@{}l@{}}Measurement\\configuration\end{tabular} & \begin{tabular}{@{}l@{}}\SI{253.7}{nm}\\probe\\laser\end{tabular} & \begin{tabular}{@{}l@{}}\SI{435.8}{nm}\\coupling\\laser\end{tabular} & \begin{tabular}{@{}l@{}}\SI{546.1}{nm}\\coupling\\laser\end{tabular} & \begin{tabular}{@{}l@{}}\SI{404.7}{nm}\\repumper\end{tabular}\\\hline\addlinespace
		(a) & \checkmark &  &  & \\\addlinespace
		(b) & \checkmark & \checkmark &  & \checkmark\\\addlinespace
		(c) & \checkmark & \checkmark & \checkmark & \\\addlinespace
		(d) & \checkmark & \checkmark & \checkmark & \checkmark\\
	\end{tabular}
	\end{ruledtabular}
	\label{tab:3PR-mconfig}
\end{table}%
The polarizations of the individual beams were chosen as given in \cite{Fry2000a}.\par
\begin{figure*}[bt]
	\centering
	\includegraphics[width=\linewidth]{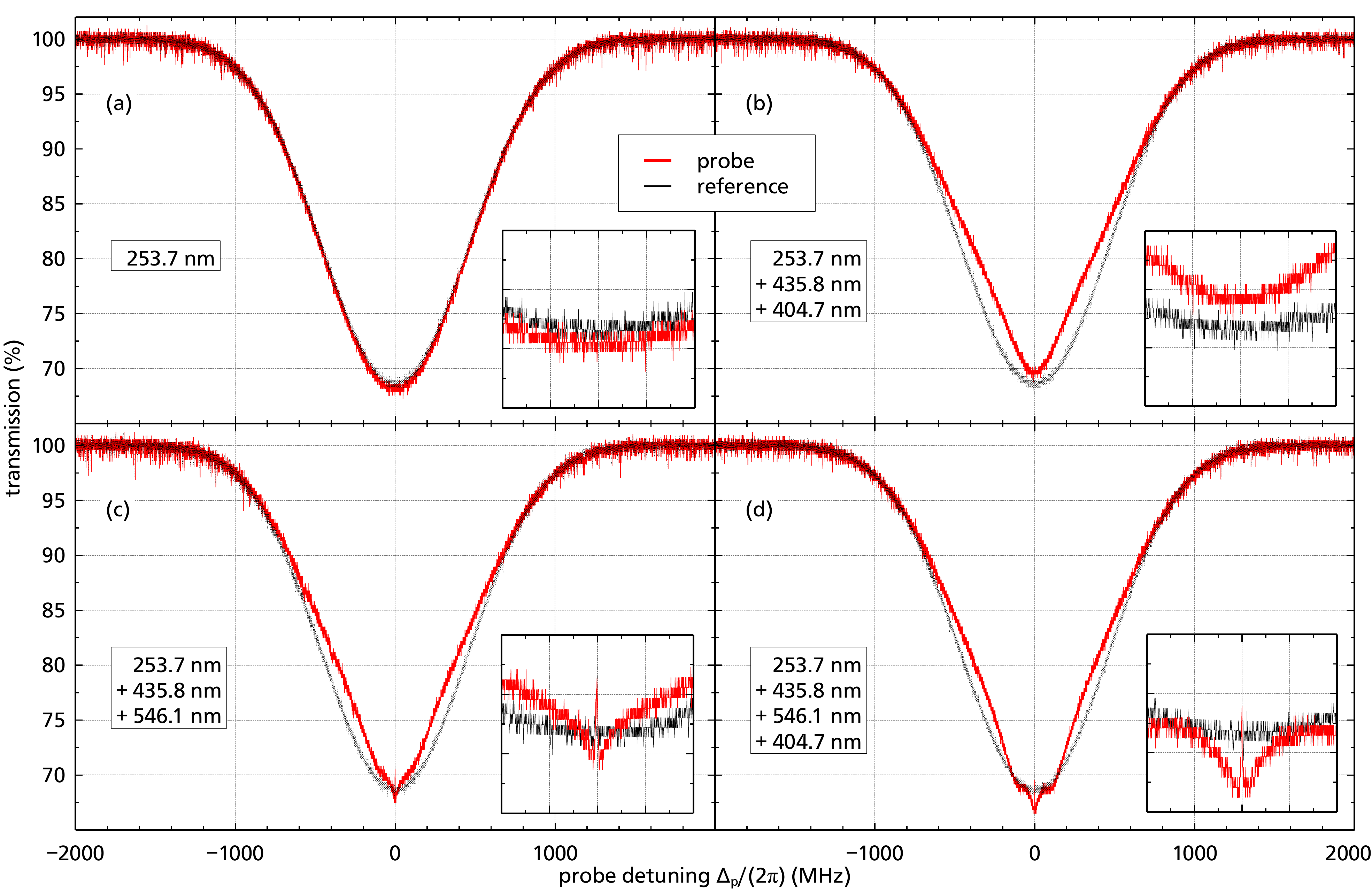}
	\caption{Normalized transmission spectra of the probe beam (red signal) and reference beam (black signal) for the four configurations in Tab. \ref{tab:3PR-mconfig}, plotted over the relative frequency detuning to the $^{202}\textrm{Hg}$ transition $\Delta_p$. The insets show a section of the transmission between \SI{66}{\%} to \SI{72}{\%} for \SI{\pm 100}{MHz} frequency detuning.}
	\label{fig:Beispielmessung-LWI-Ref}
\end{figure*}

Fig.~\ref{fig:Beispielmessung-LWI-Ref} shows the transmission of the probe beam (red signal) and reference beam (black signal) for the four measurement configurations presented in Tab. \ref{tab:3PR-mconfig}. The inset shows a zoom into the central \SI{\pm 100}{MHz} of the absorption. For frequency calibration of the x-axis a high-finesse Fabry-P\'{e}rot interferometer (FPI 100, TOPTICA Photonics AG) with a free spectral range of \SI{1.00 \pm 0.01}{GHz} was used before the second frequency doubling stage at the wavelength of \SI{507.4}{nm}.\newline
Fig.~\ref{fig:Beispielmessung-LWI-Ref} (a) corresponds to configuration (a) where probe and reference beam experience the undisturbed Doppler-broadened absorption of the $6^{1}S_{0} \leftrightarrow 6^{3}P_{1}$ transition. Considering the absorption path of \SI{2}{mm}, the measured Doppler-broadening is \SI{1013 \pm 84}{\mega \hertz}.\newline
In Fig.~\ref{fig:Beispielmessung-LWI-Ref} (b) measurement configuration (b) is applied where the $6^{3}P_{1} \leftrightarrow 7^{3}S_{1}$ transition is coherently coupled and the repumper is enabled, leading to the reduction of the probe beam's absorption, since population is pumped into the metastable $6^{3}P_{2}$ level. An EIT effect induced by the coherent coupling of the $6^{1}S_{0} \leftrightarrow 6^{3}P_{1} \leftrightarrow 7^{3}S_{1}$ ladder system is not observed, since the effective Doppler-width of the two-photon resonance of $\Delta_{\textrm{D,eff}}=\SI{445}{MHz}$ is much larger than the Rabi frequency of the coupling laser with $\Omega_{435.8} = \SI{30.8}{MHz}$ \cite{Fulton1995a,McGloin2000}.\newline
Once both coherent coupling lasers are superimposed in the UHV cell (configuration (c)), the Doppler-free three-photon coherence can develop as shown in Fig.~\ref{fig:Beispielmessung-LWI-Ref} (c). In the inset a small peak is visible, representing the three-photon coherence. As the repumper is turned off in this configuration, population is trapped in the metastable $6^{3}P_{0}$ level.\newline
In Fig.~\ref{fig:Beispielmessung-LWI-Ref} (d) all four lasers are superimposed in the UHV cell. This results in an enhanced absorption compared to the undisturbed atomic-system while the three-photon resonance is still visible, demonstrating that the additional repumper does not significantly disturb the coherent excitation.\par
\begin{figure}[htb]
	\centering
	\includegraphics[width=\linewidth]{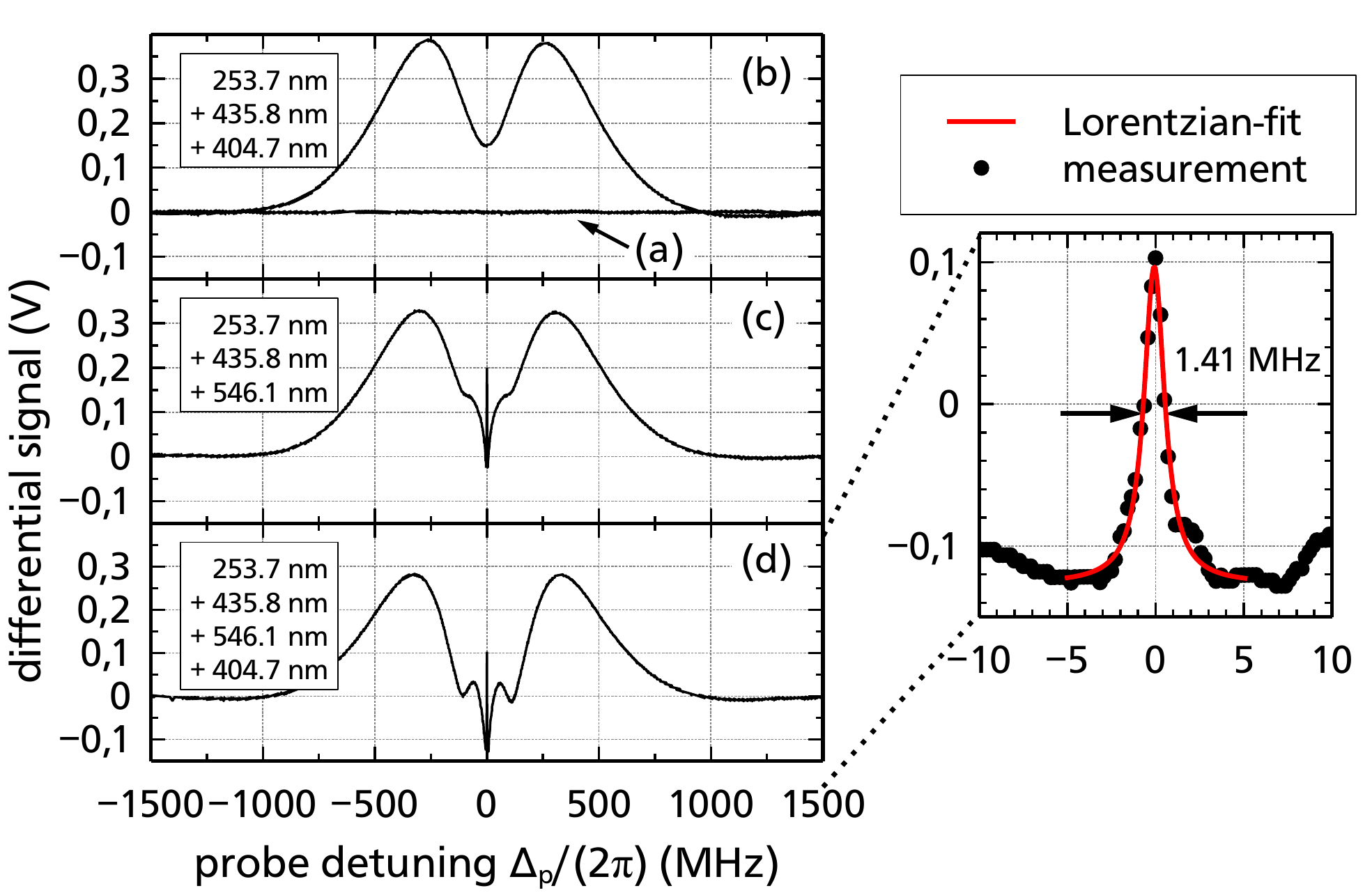}
	\caption{Difference signal of the photodiodes for the four spectra of Fig.~\ref{fig:Beispielmessung-LWI-Ref} with identical labeling of the measurements (a) to (d). A positive signal implies less absorption, while a negative signal implies more absorption compared to the undisturbed atomic system. The zoom on the right hand side shows the Doppler-free three-photon resonance peak of graph (d) with a Lorentzian fit (red line).}
	\label{fig:Beispielmessung-Diff}
\end{figure}
\begin{figure}[htb]
	\centering
	\includegraphics[width=\linewidth]{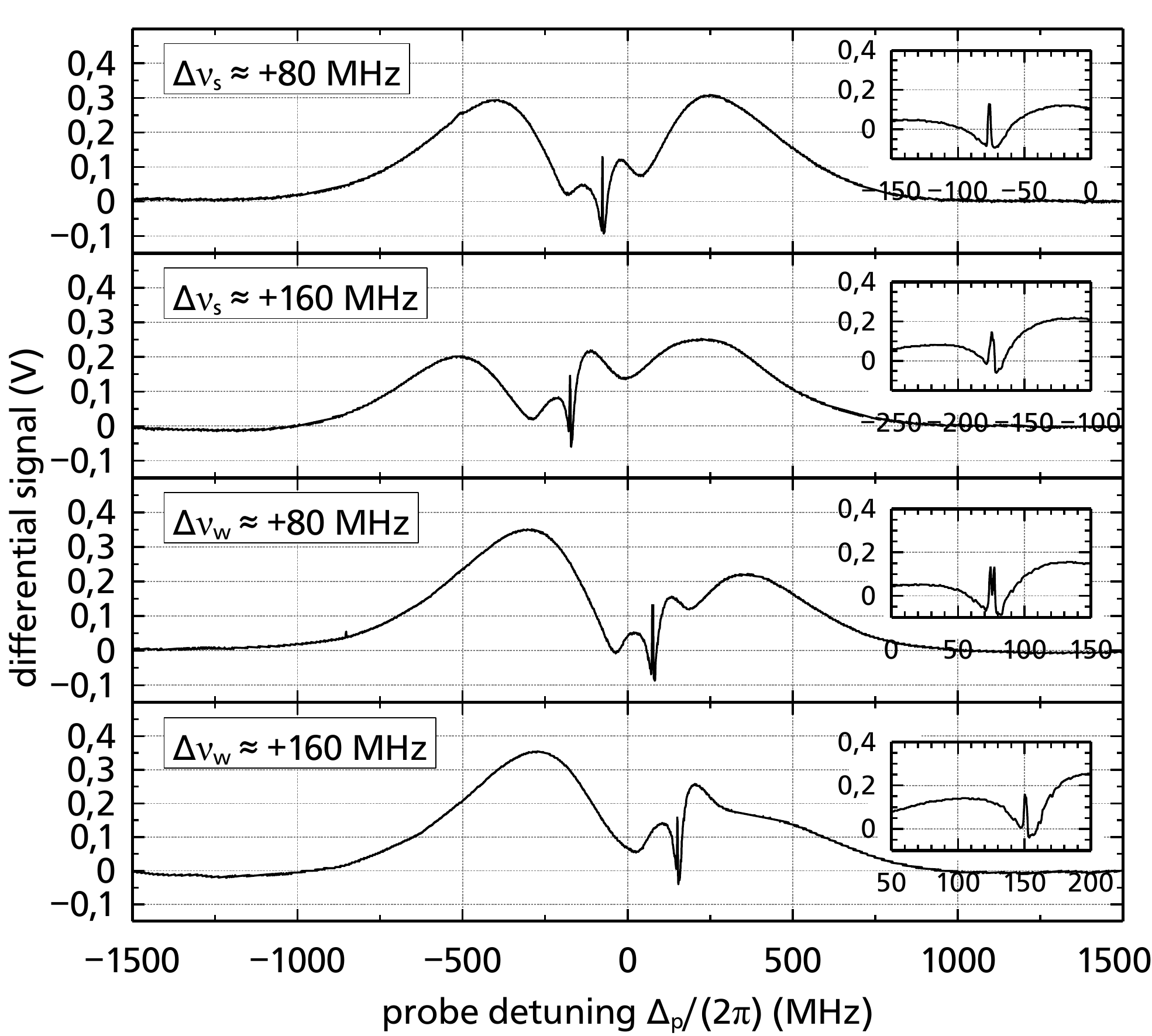}
	\caption{Difference signal for two detunings of \SI{80}{MHz} and \SI{160}{MHz} for the \SI{435.8}{nm} ($\Delta \nu_{s}$) and \SI{546.1}{nm} ($\Delta \nu_{w}$) coupling lasers, respectively.}
	\label{fig:Frequenzverschiebung}
\end{figure}

The influence of the two coupling lasers and the repumper becomes even more obvious in the difference signal compensating the residual power fluctuations of the \SI{253.7}{nm} radiation. Fig.~\ref{fig:Beispielmessung-Diff} shows the difference signal corresponding to the measurements depicted in Fig.~\ref{fig:Beispielmessung-LWI-Ref} with designations (a) to (d) being identical in both figures. While a positive signal indicates a reduction in absorption, a negative signal indicates an increased absorption compared to the undisturbed atomic system. Please note that in configuration (a) the signals are identical, which is why the difference signal is zero (configuration (a) shown within the box of part Fig.~\ref{fig:Beispielmessung-Diff} (b)) %
The difference signal of configuration (b) shows the reduction in absorption due to pumping into the metastable $6^{3}P_{2}$ level. However, for a relative frequency shift of \SI{\pm 250}{MHz} the absorption begins to increase. A probable cause is the occurrence of incoherent two-photon absorption effects such as two-step excitation (TSE)~\cite{Hayashi2010,He2014} initiating a second absorption path from the $6^{1}S_{0}$ ground state, leading to an increase in absorption.\newline
In Fig.~\ref{fig:Beispielmessung-Diff} (c) the three-photon coherence is visible as a small peak when both coupling lasers are switched on. Furthermore, an additional increase in absorption occurs in the region of \SI{\pm 50}{MHz} around the center of the transition. This effect is visible when the repumper is also enabled as shown in Fig.~\ref{fig:Beispielmessung-Diff} (d). The absorption is enhanced with respect to the undisturbed atomic system, while the three-photon resonance peak remains. The origin of the enhanced absorption is a combination of the  TSE process already mentioned, induced by the \SI{253.7}{nm} probe beam and the \SI{435.8}{nm} coupling beam, and an incoherent three-step excitation (THSE) process induced by the additional \SI{546.1}{nm} coupling beam initiating an additional absorption path from the ground state.\newline
The zoom in Fig.~\ref{fig:Beispielmessung-Diff} shows a span of \SI{\pm 10}{MHz} around the Doppler-free three-photon resonance peak of graph (d). By fitting a Lorentzian function to the peak the FWHM is determined to be \SI{1.41}{MHz}, which is slightly higher than the natural linewidth of the $6^{1}S_{0} \leftrightarrow 6^{3}P_{1}$ transition, but smaller than the expected linewidth of $\SI{1.27}{MHz} \cdot \sqrt{1 + 0.88} = \SI{1.74}{MHz}$ when saturation broadening by the probe beam is taken into account. However, the theoretical model (cf. Sec. \ref{sec:theory}) predicts a FWHM of \SI{704}{kHz}, which is half of the measured width. The broadening is due to the remaining frequency fluctuations of the two coherent coupling lasers. For the Doppler-free angle configuration, the shift of the three-photon resonance is still dependent on the frequency shifts of the two coupling lasers. This is shown in Fig.~\ref{fig:Frequenzverschiebung} where in the first two graphs the \SI{435.8}{nm} coupling laser is frequency shifted by $\Delta \nu_{s}$=\SI{+80}{MHz} and \SI{+160}{MHz} and in the last two graphs the \SI{546.1}{nm} coupling laser is frequency shifted by $\Delta \nu_{w}$=\SI{+80}{MHz} and \SI{+160}{MHz} with respect to their atomic transition. %
The three-photon resonance is shifted to negative frequencies for a positive detuning of the \SI{435.8}{nm} laser and to positive frequencies for a positive detuning of the \SI{546.1}{nm} laser. This is the expected behavior according to Eqn.~\ref{equ:Doppler-broadening}. This in turn means that every frequency deviation of the two coupling lasers during the scan of the \SI{253.7}{nm} probe laser results in a broadening of the three-photon resonance peak. For the measurements in Fig.~\ref{fig:Frequenzverschiebung} the detuned laser could not be frequency stabilized as it was too far from resonance. This leads to even broader three-photon resonance peaks and even a double structure induced by a frequency hop of the \SI{546.1}{nm} laser which is shown in the inset for $\Delta \nu_{546.1} = \SI{80}{MHz}$. Furthermore, the three-photon resonance peak becomes more and more asymmetric with increasing frequency detuning of the coupling lasers, as shown in the insets of Fig.~\ref{fig:Frequenzverschiebung}. This behavior is similar to that of three-level EIT systems where the coupling laser is frequency detuned and is based on the asymmetry of the Autler-Townes splitting which results in dispersive like structures~\cite{Li1995,Kim2001a,Bhattacharyya2004,Moon2014a}. \par
\begin{figure}[htb]
	\centering
	\includegraphics[width=\linewidth]{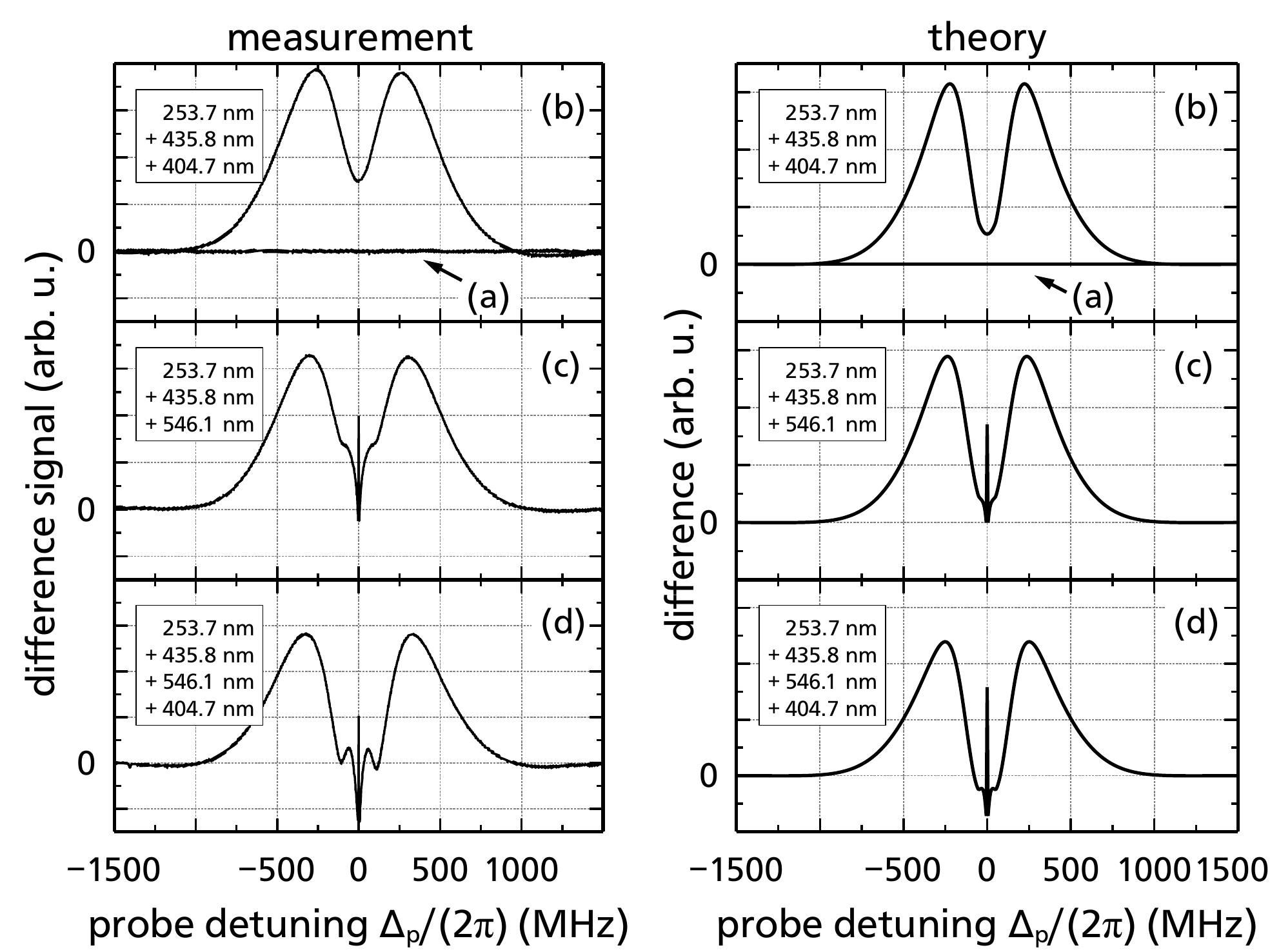}
	\caption{Comparison of the difference signals for the measurement (left hand side) and the simulation (right hand side). The letters (a) to (d) correspond to the experimental configurations discussed in table~\ref{tab:3PR-mconfig}.}
	\label{fig:Vergleich-Messung-Sim}
\end{figure}

Fig.~\ref{fig:Vergleich-Messung-Sim} shows a comparison of the measurements of Fig.~\ref{fig:Beispielmessung-Diff} and simulated spectra calculated with the theoretical model presented in~\cite{Sturm2014}. %
For the simulation, parameters equivalent to the experimental situation were used (e.g. Rabi frequencies, temperature of the mercury atoms, absorption path). The simulation shows the same processes, such as TPA, which could be observed in the measurement but the strength and width of the structures differ between simulation and measurement. One cause is probably the beam profiles which are taken to have a uniform intensity distribution in the simulation, but resemble a Gaussian intensity distribution in the experiment. This leads to a uniform overlap region of the laser beams in the simulation while for the experiment the overlap region possesses a complex spatial distribution of the coupling laser intensities which could lead to a broadening of the measured structures. 
%
%
\section{Possibility of AWI in mercury}
\begin{figure}[htb]
	\centering
	\includegraphics[width=\linewidth]{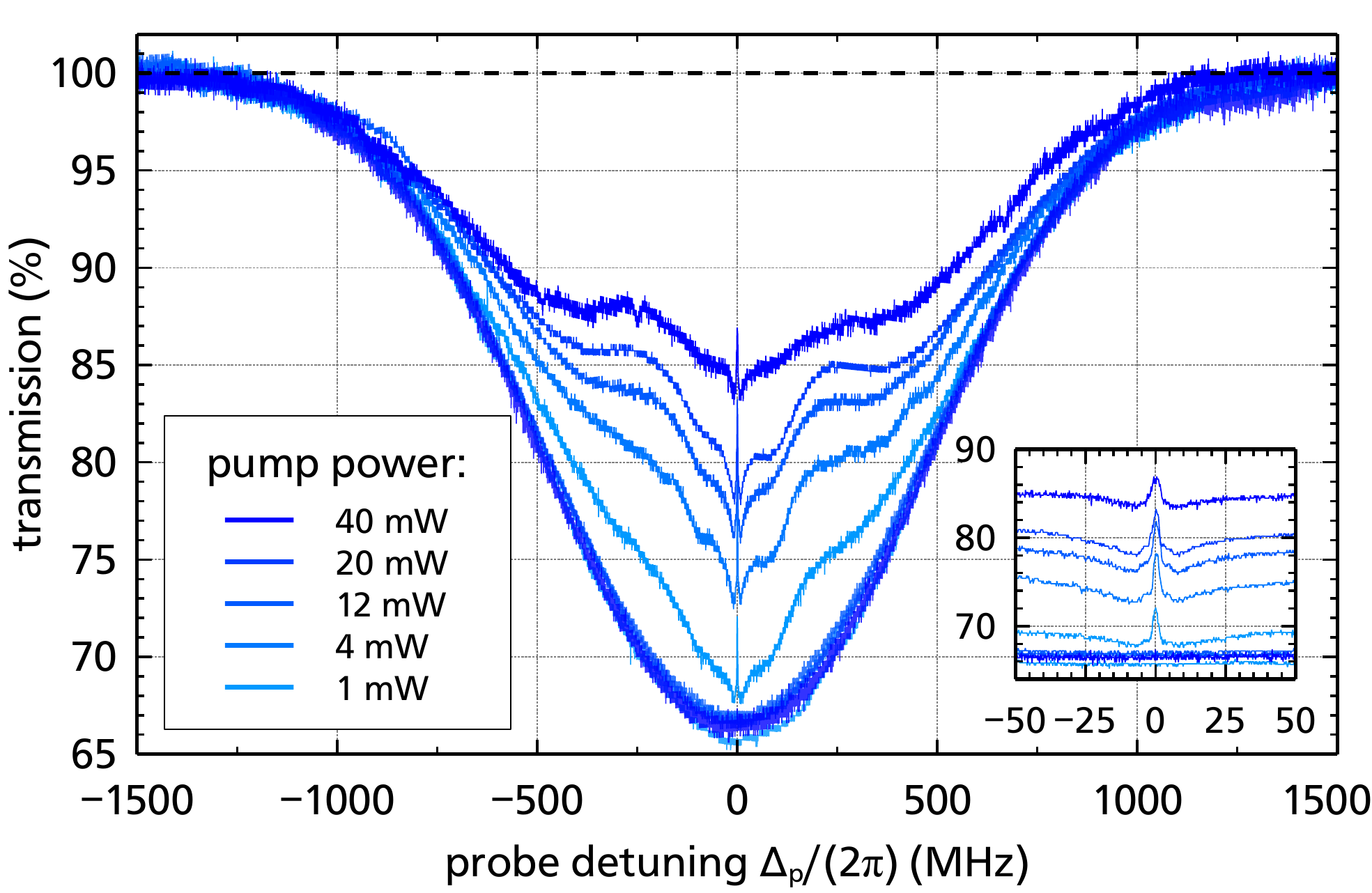}
	\caption{Transmission of the probe beam for the AWI configuration and various pump powers. For each measurement the transmission for the undisturbed UV transition is plotted in addition. The inset shows a zoom of the central \SI{\pm 50}{MHz}.}
	\label{fig:AWI-Transmissionspektren}
\end{figure}

Measuring the Doppler-free three-photon coherence is the first step towards LWI in mercury as proposed in \cite{Fry2000a}. The next step is to use an additional incoherent pump on the $6^{1}S_{0} \leftrightarrow 6^{3}P_{1}$ transition which corresponds to an AWI scheme and is implemented as shown in Fig.~\ref{fig:UV-Strahlengang}.\newline
The angle between the incoherent pump beam and the probe beam measures \SI{5}{\degree} so that both beams are well separated after traveling through the UHV cell (cf. Fig.~\ref{fig:UV-Strahlengang}). To achieve a good overlap with the probe beam within the UHV cell, the beam profile of the pump beam exhibits an elliptical shape of \SI{2.5}{mm} for the long axis which is parallel to the optical table and \SI{1.4}{mm} for the short axis. The polarization of the pump beam is linear and parallel to the probe beam's linear polarization.\newline
Fig.~\ref{fig:AWI-Transmissionspektren} shows the transmission of the probe beam in measurement configuration (d) with pump beam for five different pump powers from \SI{1}{mW} to \SI{40}{mW}. The absorption spectrum of the undisturbed UV transition is shown in addition for each measurement.

The power of the probe  and reference beams was reduced to \SI{15}{\micro W} for this measurement to decrease the saturation of the $6^{1}S_{0} \leftrightarrow 6^{3}P_{1}$ transition. This power corresponds to a Rabi frequency of $\Omega = 2\pi \cdot \SI{0.175}{MHz}$ and a saturation parameter of $S_0 = 0.27$.

\begin{figure}[htb]
	\centering
	\includegraphics[width=\linewidth]{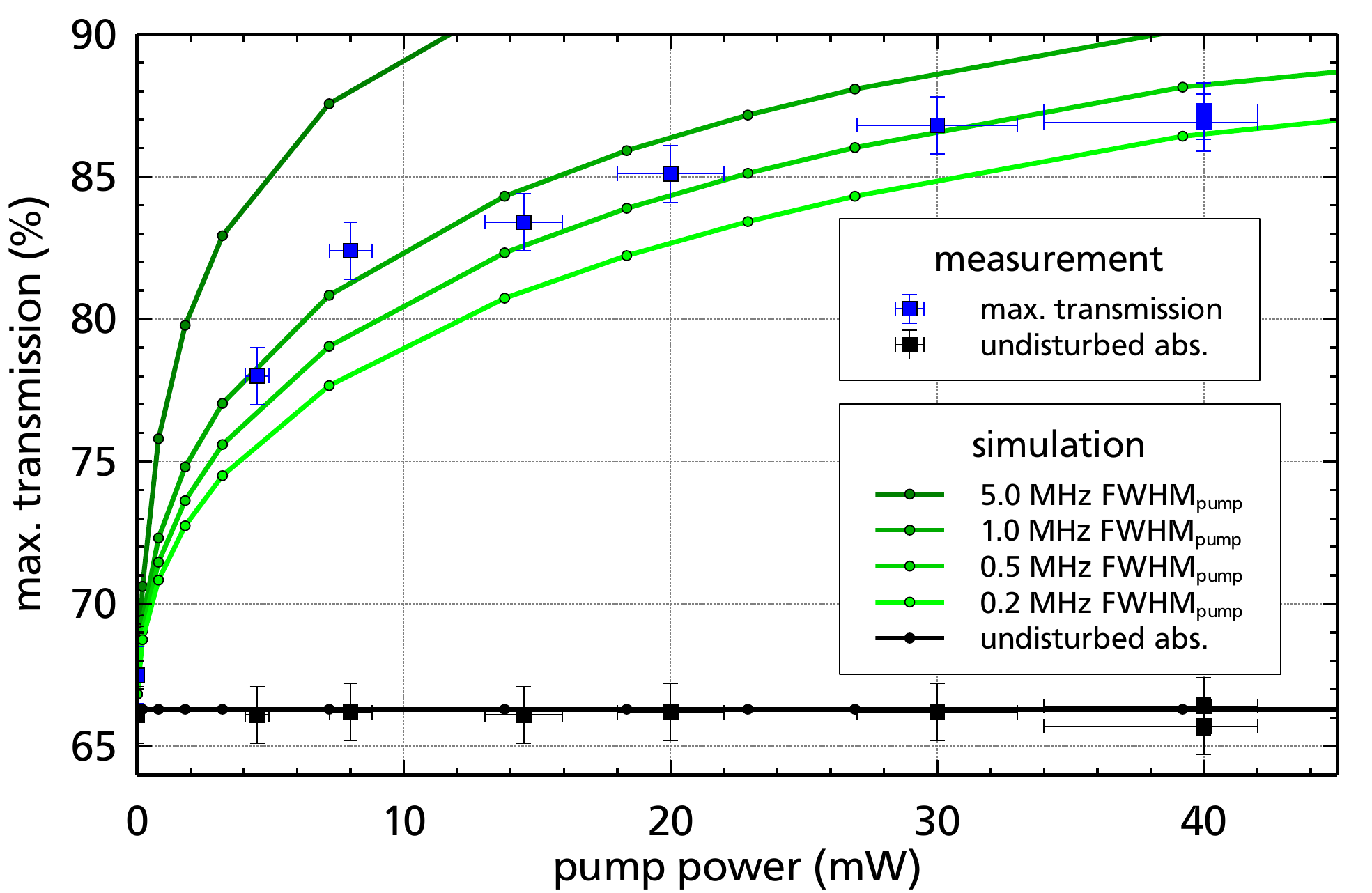}
	\caption{Measured maximum transmission of the Doppler-free three-photon resonance peak in relation to the pump power (blue squares). The absorption of the undisturbed Doppler-broadened transition are depicted as black squares. The calculated values from the theoretical model are plotted as green lines for four different spectral width of the pump $\textrm{FWHM}_{\textrm{pump}}$.}
	\label{fig:AWI-Transmission-Simulation}
\end{figure}

\begin{figure}[htb]
	\centering
	\includegraphics[width=\linewidth]{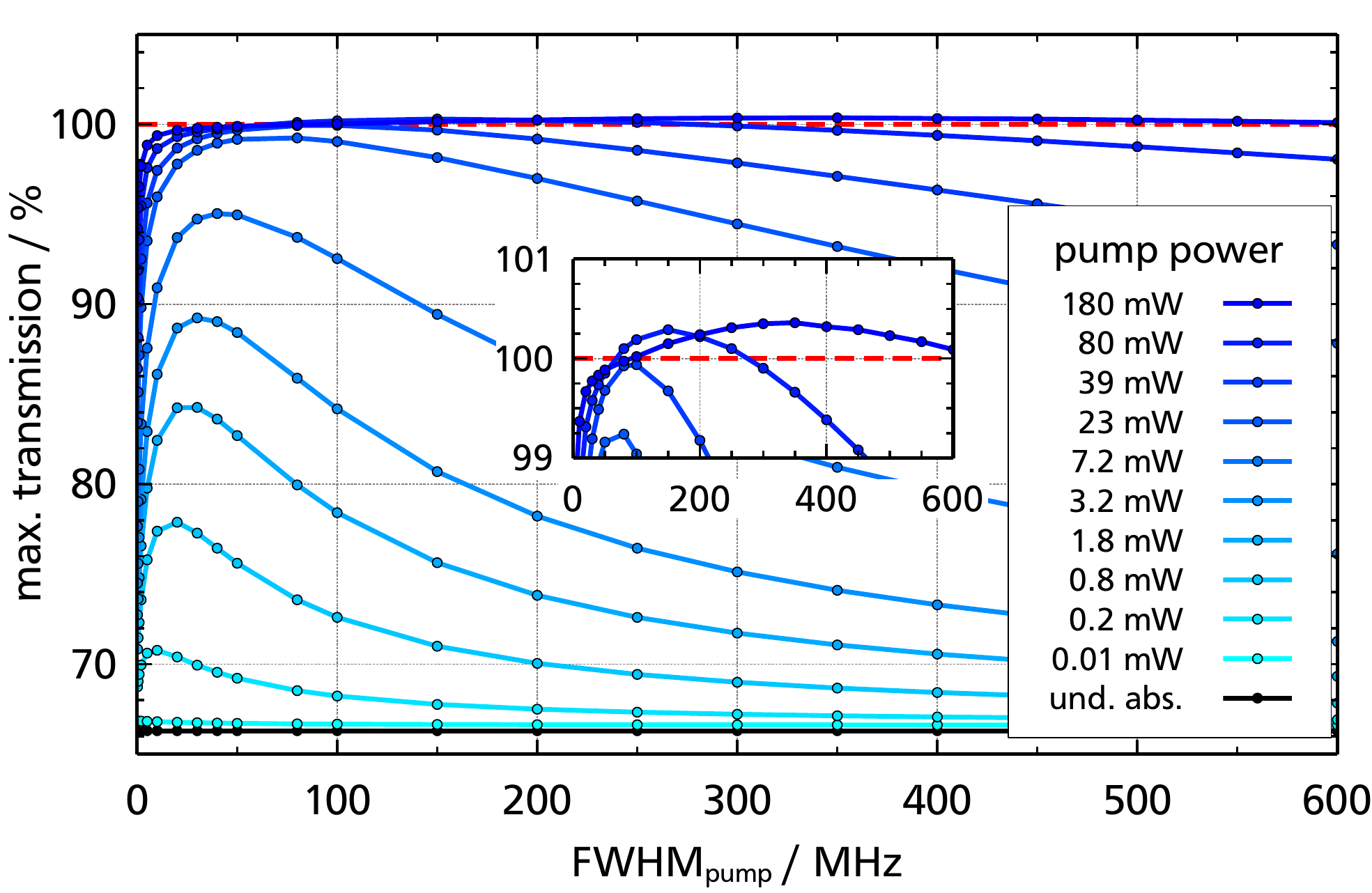}
	\caption{Calculated maximum transmission of the Doppler-free three-photon resonance peak in relation to the spectral width of the pump $\textrm{FWHM}_{\textrm{pump}}$ for different pump powers. The inset shows the region where AWI should be possible.}
	\label{fig:Sim-Transmission}
\end{figure}

The achieved maximum transmission of the three-photon resonance is \SI{86.9}{\%} for a pump power of \SI{40}{mW} and is mainly limited by the insufficient spectral width of the pump of  about \SI{0.5}{MHz} 
(cf.~Sec.~\ref{sec:set-up}). This is shown in Fig.~\ref{fig:AWI-Transmission-Simulation} where the maximum transition of the three-photon resonance peak is plotted against the pump power and compared to simulated data for four different widths of the pump. %
The simulated data show a strong dependence on the pump width, which becomes even more obvious in Fig.~\ref{fig:Sim-Transmission} where the simulated maximum transition of the three-photon resonance peak is plotted against the pump width for different pump powers. %
The strong dependence on the pump width is based on the Doppler-free nature of the three-photon coherence. Because it is Doppler-free, all mercury atoms independent of their velocity contribute to the three-photon coherence. However, as the pump width is very small, it could only address a small part of the Doppler-broadened spectrum. The broader the pump width, the more atoms could be addressed, hence increasing the pump efficiency.\newline
The simulation predicts a certain spectral width for every pump power where the pump efficiency reaches its maximum. As the pump width is further increased, for a constant pump power, the spectral intensity decreases and the pump efficiency begins to decrease.\newline
The inset in Fig.~\ref{fig:Sim-Transmission} shows the region where a transmission $> \SI{100}{\%}$ could be achieved. AWI is predicted for a pump power $> \SI{80}{mW}$ and a spectral width $\textrm{FWHM}_\textrm{pump} \gtrsim \SI{70}{MHz}$.\newline
While in principle it is possible to achieve a pump power of \SI{80}{mW} with the existing \SI{253.7}{nm} laser system, the pump width is limited by the AOM driver employed to \SI{0.5}{MHz}. To address the whole bandwidth of the AOM, it is necessary to build a discrete AOM driver which is able to transfer the white noise modulation to the AOMs carrier frequency. This approach is planned in future experiments.%
%
%
\section{Conclusion}
In conclusion, the first measurements of a Doppler-free three-photon coherence based on a DDKS in mercury vapor were presented. Because of the large differences of the involved wavelengths of \SI{253.7}{nm}, \SI{435.8}{nm} and \SI{546.1}{nm}, a special geometrical configuration is necessary to compensate the Doppler-shifts in the thermal mercury vapor. Such a geometrical configuration was also investigated in \cite{Ye2002} but without any effect because of the experimental setup.\newline
The three-photon coherence is the basis of a LWI scheme facilitating LWI at a wavelength of \SI{253.7}{nm} and whose feasibility was shown by detailed calculations in \cite{Fry2000a,Sturm2014} (cf. Sec. \ref{sec:theory}). There are also considerations using a three-photon resonance between $6^{1}S_{0} \leftrightarrow 6^{3}P_{0}$ states for an optical clock \cite{Hong2005}, which should also be possible with this system by exchanging the roles of the \SI{546.1}{nm} and \SI{404.7}{nm} lasers.\newline
Using different laser configurations, the influences of the coupling lasers and the repumper were investigated and it was shown that the coherence only occurs when both coupling lasers are enabled. By shifting the coupling lasers away from the atomic resonance, the three-photon coherence experiences a corresponding shift as predicted by the theory. This in turn is the cause for the larger width of the three-photon resonance peak compared to the simulation. \newline
An AWI scheme was implemented and the measurement results were compared to the simulation. It became apparent that the linewidth of the \SI{253.7}{nm} pump is a crucial factor, since atoms of all velocity classes  contribute to the three-photon coherence. On the basis of these results, it was possible to predict AWI for a pump power $> \SI{80}{mW}$ and a spectral width $\gtrsim \SI{70}{MHz}$, which would be the basis for the first LWI measurements with a wavelength in the UV.
%
%
%
\begin{acknowledgments}
The authors thank the Deutsche Forschungsgemeinschaft (DFG) under Grant WA 1658/2-1 for support of this research. M.R.S. and R.W. acknowledge support from the German Aeronautics and Space Administration (DLR)
through Grant 50 WM 1557. We gratefully acknowledge Holger John for providing the radiation of his \SI{507.4}{nm} laser system. 
\end{acknowledgments}

\appendix*

\section{Bloch equations}
The coupling of the atoms to the electromagnetic field induces coherent dynamics described by the Hamilton operator in Eq.~(\ref{equ:4NS-Hamiltonian}) as well as the incoherent radiation damping. This is described the optical Bloch equations
\begin{equation}
\partial_t \hat{\rho} = \left( \mathcal{L}_\text{c} + \mathcal{L}_\text{i} \right) \hat{\rho}.
\label{eqn:Mastereqn}
\end{equation}
for the atomic density operator $\hat{\rho}$. Within the Born-Markov approximation, one finds
\begin{equation}
\mathcal{L}_\text{c} \hat{\rho} = -\frac{i}{\hbar} \big[ \hat{H},\hat{\rho} \big]
\end{equation}
for the coherent evolution and 
\begin{equation}
\begin{split}
\mathcal{L}_\text{i} \hat{\rho} = &\sum\limits_{j \in \mathcal{T}}
	\Gamma_j \frac{n_j+1}{2} \left( 2 \hat{s}_j \hat{\rho} \hat{s}_j^\dagger - \hat{s}_j^\dagger \hat{s}_j \hat{\rho}	-\hat{\rho} \hat{s}_j^\dagger \hat{s}_j\right) \\
	&+ \sum\limits_{j \in \mathcal{T}} \Gamma_j \frac{n_j}{2} \left( 2 \hat{s}_j^\dagger \hat{\rho} \hat{s}_j - \hat{s}_j \hat{s}_j^\dagger \hat{\rho}-\hat{\rho} \hat{s}_j \hat{s}_j^\dagger\right).
\end{split}
\end{equation}
for the incoherent dynamics.
Here, $\mathcal{T} = \{ 12, 23, 34 \}$ is the set of atomic transitions whereas $n_j$ and $\hat{s}_j$ denote the corresponding mean photon numbers and lowering operators. The thermal population of optical modes at room temperature is negligible, i.e. $n_j=0$. However, we model an incoherent pump on the $1 \leftrightarrow 2$ transition
by setting $n_{12}=r/\Gamma_{12}$ proportional to the pump rate $r$.

For the  four-level system considered here, we obtain the following set of Bloch equations for the density matrix
\begin{widetext}
\begin{align}
	\dot{\rho}_{11} =& 
	- r \rho_{11} + 
	(\Gamma_{12}+r) \rho _{22}
	- i  \Omega_p ^*  \rho_{12}
	+i  \Omega_p  \rho_{12}^* 
	\\
	\dot{\rho}_{22} =& 
	- (\Gamma_{12}+r)  \rho _{22} 
	+ \Gamma_{23} \rho_{33}
		+ r  \rho_{11}
	+ i  \Omega_p^*  \rho_{12} 
	- i  \Omega_p\rho_{12}^* 
	- i  \Omega_s^*  \rho _{23} 
	+ i \, \Omega_s \, \rho_{23}^*  
	\\
	\dot{\rho}_{33} =& 
	- (\Gamma_{23}+\Gamma_{34}) \rho _{33}
	+i  \Omega_s^* \rho _{23}
	-i  \Omega_s \rho _{23}^*
	+i  	\Omega_w^*  \rho_{34}^*
	- i  \Omega_w \rho _{34} 
	\\
	\dot{\rho}_{44} =& +
	  \Gamma_{34}  \rho _{33}
	-i  \Omega_w ^*  \rho_{34}^*
	+i  \Omega_w  \rho _{34} 
	\\
	\dot{\rho}_{12} =& 
	- \left[ i  \Delta_p +\Gamma_{12}/2 + r\right] \rho_{12}
	- i  \Omega_s^*  \rho_{13}
	+i \Omega_p \left(\rho _{22}-\rho _{11}\right)
	\\
	\dot{\rho}_{13} =& 
	-\left[ i \left(\Delta_p +\Delta_s\right)
	+\left( \Gamma_{23}+\Gamma_{34}+r\right)/2\right] \rho_{13} 
	+ i \Omega_p \rho _{23}
	-i  \Omega_s  \rho_{12} 
	- i  \Omega_w  \rho_{14}
	\\
	\dot{\rho}_{14} =& 
	- \left[i\left(\Delta_p + \Delta_s-\Delta_w\right) +r/2\right] \rho _{14}
	-i  \Omega_w ^*  \rho _{13} 
	+ i  \Omega_p  \rho_{24} 
	\\
	\dot{\rho}_{23} =& 
	-\left[i \Delta_s + \left(\Gamma_{12}+\Gamma_{23}+\Gamma_{34}+ r \right)/2 \right] \rho_{23}
	+ i  \Omega_p ^* \rho_{13}
	- i  \Omega_s  \rho_{22} 
	+ i  \Omega_s  \rho_{33}
	-i\Omega_w  \rho_{24}
	\\
	\dot{\rho}_{24} =& 
	-\left[i \left(\Delta_s-\Delta_w\right)+\left( \Gamma_{12} + r \right) /2 \right] \rho_{24} 
	+i \Omega_p^*  \rho_{14} 
	+ i  \Omega_s  \rho _{34}
	-i \Omega_w ^*  \rho _{23}
	\\
	\dot{\rho}_{34} =&  
	-\left[-i \Delta_w+\left(\Gamma_{23}+\Gamma_{34} \right)/2\right]\rho _{34}
	+ i  \Omega_s ^* \rho _{24} 
	+ i  \Omega_w ^* \left(\rho_{44} - \rho_{33}\right)
\end{align}
\end{widetext}
From the stationary state $\dot{\rho}=0$ of the Bloch-equations, subject to probability conservation
$\textrm{Tr}[\rho]
=1$, we obtain the linear susceptibility on the 
$1 \leftrightarrow 2$ transition
\begin{equation}
\chi^{(1)} = \chi' + i \chi'' = \frac{\left| d_{21} \right|^2 \mathcal{N}}{\epsilon_0 \hbar \Omega_p^*} \rho_{21}.
\end{equation}


\bibliography{Three-photon-coherence}

\end{document}